\documentclass[summary]{ursi}
\usepackage[bookmarks=false]{hyperref}
\usepackage{graphicx,color,epsfig}
\usepackage{amssymb,epsf,amsmath}

\newcommand{\beq}{\begin{equation}}
\newcommand{\eeq}{\end{equation}}
\newcommand{\beqn}{\begin{eqnarray}}
\newcommand{\eeqn}{\end{eqnarray}}

\title{Explainable machine learning workflows for radio astronomical data processing}

\author{S. Yatawatta, A. Ahmadi, B. Asabere, M. Iacobelli, N. Peters, M. Veldhuis*
  }

\affiliation{%
  ASTRON, The Netherlands Institute for Radio Astronomy, Dwingeloo, The Netherlands, https://www.astron.nl/
}

\begin{document}

\maketitle
\begin{abstract}
  Radio astronomy relies heavily on efficient and accurate processing pipelines to deliver science ready data. With the increasing data flow of modern radio telescopes, manual configuration of such data processing pipelines is infeasible. Machine learning (ML) is already emerging as a viable solution for automating data processing pipelines. However, almost all existing ML enabled pipelines are of black-box type, where the decisions made by the automating agents are not easily deciphered by astronomers. In order to improve the explainability of the ML aided data processing pipelines in radio astronomy, we propose the joint use of fuzzy rule based inference and deep learning. We consider one application in radio astronomy, i.e., calibration, to showcase the proposed approach of ML aided decision making using a Takagi-Sugeno-Kang (TSK) fuzzy system. We provide results based on simulations to illustrate the increased explainability of the proposed approach, not compromising on the quality or accuracy.
\end{abstract}

\section{Introduction}
Numerous data processing steps are required by radio telescopes to produce science ready results from raw observations. In order to cope with large volumes of data, such processing workflows need to operate in almost real time, with highest computational efficiency, but without compromising the quality of the data. To achieve these goals, most of these pipelines are configured and tuned by expert astronomers although ML is becoming a viable alternative \cite{Y2021,YRL2023}. A typical ML model that is trained for a specific task appears as a a black-box to the end user. This is a hindrance for building confidence in their use in data processing. Furthermore, enhancing such models with the existing human expertise requires a mechanism for understanding the behavior of such black-box ML models.

\begin{figure}[ht]
  \begin{minipage}{0.99\linewidth}
    \begin{center}
  \epsfig{figure=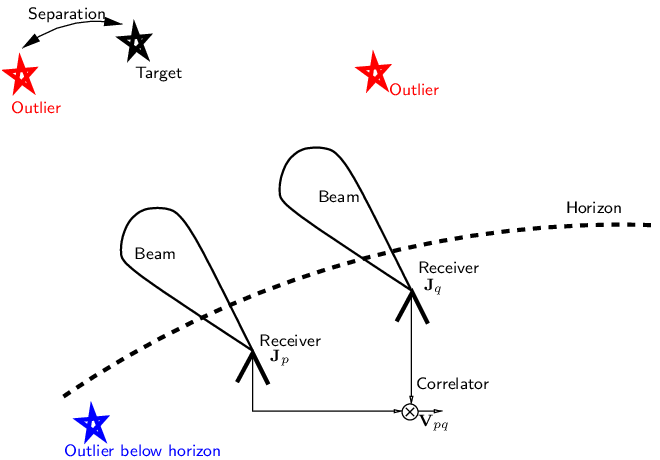,width=8.0cm}\\
    \end{center}
  \end{minipage}
  \caption{A geometrical overview of the observing setup. While a target area in the sky is being observed, several strong outlier sources act as interference.\label{geometry}}
\end{figure}

In this paper we propose the use of fuzzy inference \cite{TSK,ANFIS} to improve the explainability of traditional ML models (see \cite{Zhen2022} for an in-depth survey). We select direction dependent calibration for outlier source removal to apply the proposed technique. As shown in Fig. \ref{geometry}, bright outlier sources in the sky act as interference to a target being observed, especially in low frequency observations (LOFAR \cite{LOFAR}). Making the correct choice of which outliers to remove from the data is a matter of importance (in addition to other hyperparameters such as the time-frequency resolution of the calibration). It is possible to use a purely data-driven approach to make this choice \cite{yatawatta2023hint} but under noisy conditions (very low frequency observations) this becomes unreliable. Note also that selecting all the outliers for removal is also not feasible -- because increasing of the number of free parameters (to solve for) will make the problem ill-posed. Therefore, the optimal choice of the outlier selection for removal needs to be made uniquely for each observation. Noting the astrometric stability of the positions of the outliers with respect to a target in the sky, we propose a model-driven approach, ideally suitable for the use of ML methods. In order to make the ML models more explainable, we also propose the use of Fuzzy inference as an addition to traditional multilayer perceptron based ML models.

The rest of the paper is organized as follows: in section \ref{sec:cal}, we describe  directional dependent calibration for outlier removal and the data-driven approach for making the optimal choice. In section \ref{sec:fuzzy}, we provide a brief overview of fuzzy inference. Next in section \ref{sec:results}, we provide results (based on simulations) for the feasibility of the proposed model-driven approach as well as the explainability of the trained models and finally in section \ref{sec:conclusions}, we draw our conclusions.

\section{Directional calibration for outlier removal\label{sec:cal}}
Given a pair of receivers $p$ and $q$ (out of a total of $N$) on Earth as in Fig. \ref{geometry}, the output at the correlator can be given as \cite{HBS}
\beq \label{vpq}
{\bf v}_{pq}=\sum_{i=0}^K {\bf s}_{pqi} +{\bf n}_{pq}
\eeq
where ${\bf s}_{pqi}$ is the signal from the $i$-th direction in the sky ($K$ outliers) and ${\bf n}_{pq}$ is the noise \cite{yatawatta2023hint}. We stack up ${\bf v}_{pq}$ in (\ref{vpq}) for all possible $p,q$ and for all time and frequency ranges within which a single solution for calibration is obtained (say ${\bf y}$). We can do the same for the right hand side of (\ref{vpq}) for each direction $i$ to get ${\bf s}_i$ and also for the noise to get ${\bf n}$. Let us rewrite (\ref{vpq}) after stacking as
\beq \label{cal}
{\bf y}={\bf s}_0+\sum_{i\in \mathcal{I}} {\bf s}_{i} +{\bf n} 
\eeq
where we have ${\bf s}_0$ for the model of the target direction and the remaining directions are taken from the set $\mathcal{I}$. After calibration, we find the residual signal where the outlier sources have been removed as
\beq \label{res}
{\bf r}={\bf y}-\widehat{\bf s}_0-\sum_{i\in \mathcal{I}} \widehat{\bf s}_{i}
\eeq
where $\widehat{\bf s}_i$ is the model constructed by calibration (so not necessarily the true model). The problem we need to tackle is to determine $\mathcal{I}$ out of all possible outlier sources. A data-driven approach would use (\ref{res}), and find the Akaike information criterion (AIC \cite{Akaike}) as
\beq \label{AIC}
AIC_{\mathcal{I}}=\left(\frac{N \sigma_{\bf r}}{\sigma_{\bf y}}\right)^2+N|\mathcal{I}|
\eeq
where $\sigma_{\bf y}$ and $\sigma_{\bf r}$ are the standard deviations of ${\bf y}$ and ${\bf r}$ in (\ref{res}), respectively. The cardinality of $\mathcal{I}$ is given by $|\mathcal{I}|$ (proportional to the number of free parameters). By trying out all possible choices for $\mathcal{I}$, we select the outlier directions that minimize the AIC as the optimal choice. However, under very noisy conditions, the ratio $\frac{\sigma_{\bf r}}{\sigma_{\bf y}}$ in (\ref{AIC}) will hardly change for the various choices of $\mathcal{I}$, making the data-driven approach less reliable and motivating us to select a model-driven approach.

\section{Fuzzy inference\label{sec:fuzzy}}
A fuzzy variable is more related to a concept or a linguistic (semantic) value rather than an exact mathematical number \cite{TSK,ANFIS}.
Consider a vector of (crisp or exact)  input values ${\bf x}$ of size $D\times 1$, where $x_d$ is the $d$-th input feature. The antecedent and the consequent of a TSK fuzzy system with $R$ rules (first order) can be described as
\beqn \label{fuzz}
\mathrm{Rule}_r:\ \mathrm{IF}\ x_1\ \mathrm{is}\ X_{r,1}\ \mathrm{and}\ x_2\ \mathrm{is}\ X_{r,2}\  \ldots\ \mathrm{and}\ x_D\ \mathrm{is}\ X_{r,D}\\\nonumber
\mathrm{THEN}\ y = w_1 x_1+ w_2 x_2 +\ldots + w_D x_D + b
\eeqn
where $X_{r,d}$ is the fuzzy membership function of the $d$-th input feature in the $r$-th rule. The crisp output $y$ of the consequent is determined by the weights $w_d$ and the bias $b$. The weights are determined by the degree of membership of the antecedent (and additional linear transforms if necessary).
For our example, we select Gaussian membership functions to represent $X_{r,d}$ in (\ref{fuzz}) that are described as
\beq\label{gmf}
\mu_{r,d}(x_d)=\exp \left(-\frac{(x_d-m_{r,d})^2}{2\sigma_{r,d}^2}\right)
\eeq
where $m_{r,d}$ and $\sigma_{r,d}^2$ denote its mean and the variance.

The degree of fulfillment (or the firing rate) of each rule $f_r({\bf x})$ can be calculated in many ways \cite{Gu2020} and we use the product 
\beq\label{fr}
f_r({\bf x}) =\Pi_{d=1}^{D} \mu_{r,d}(x_d).
\eeq

In order to calculate the output (defuzzification) weights $w_d$, we use
\beq\label{defuzz}
\overline{f_r({\bf x})} =\frac{f_r({\bf x})}{\sum_{i=1}^{R} f_i({\bf x})}
\eeq
which is similar to the {\em SoftMax} operation in deep learning. Additional multilayer perceptrons can be added to the outputs of (\ref{defuzz}) before calculating the crisp output \cite{Gu2020,Cui2021}.
\section{Results\label{sec:results}}
\begin{figure}[htbp]
  \begin{minipage}{0.99\linewidth}
\begin{center}
  \begin{minipage}{0.99\linewidth}
\centering
  \centerline{\includegraphics[width=0.7\textwidth]{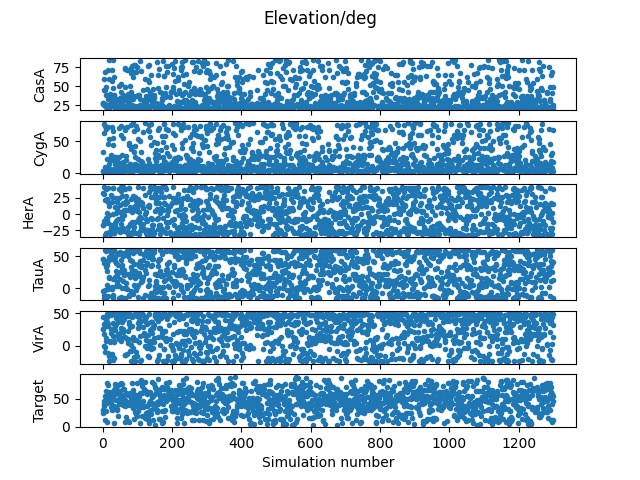}}
\vspace{0.01cm} \centerline{(a)}
\end{minipage}
  \begin{minipage}{0.99\linewidth}
\centering
  \centerline{\includegraphics[width=0.7\textwidth]{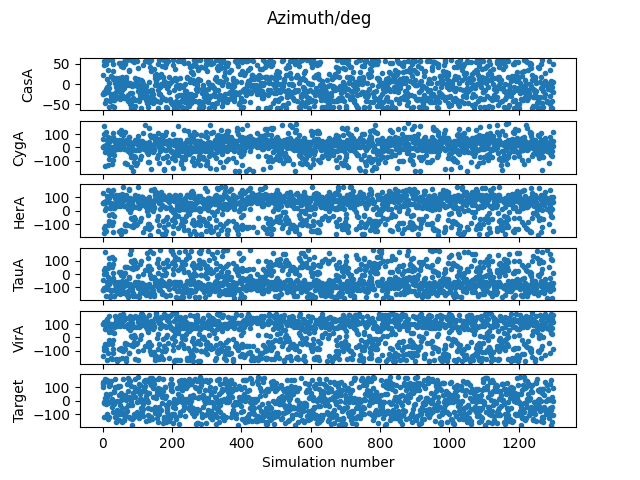}}
\vspace{0.01cm} \centerline{(b)}
\end{minipage}
  \begin{minipage}{0.99\linewidth}
\centering
  \centerline{\includegraphics[width=0.7\textwidth]{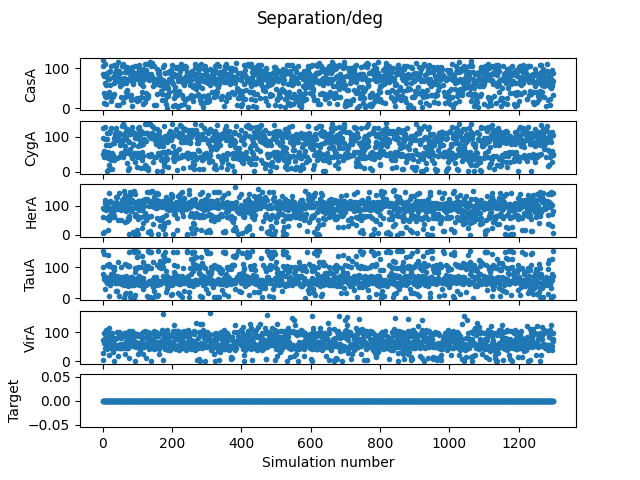}}
\vspace{0.01cm} \centerline{(c)}
\end{minipage}
\end{center}
\end{minipage}
  \caption{The statistical spread of the simulated data used in training the ML model: (a) elevation (b) azimuth (c) separation. \label{data_dist}}
\end{figure}

We simulate data for the LOFAR telescope for various configurations as listed in Table \ref{configs}. The target direction is randomly selected to be anywhere in the sky and the epoch of the observation is also randomly selected (provided the target is above the horizon).
\begin{table}[htbp]
  \centering
  \caption{The simulation and deep learning settings\label{configs}}
\begin{tabular}{l|c|}
  Outliers & CasA, CygA, TauA,\\
  {} & VirA, HerA\\
  Number of outliers $K$ & 5\\
  Observation duration & 30 s\\
  Number of subbands & 3\\
  Frequency range & $[15,70]$ or $[110,180]$ MHz\\
  Stations $N$ & 46 (LBA) or 62 (HBA)\\
  Signal to noise ratio & {}\\
  $\|{\bf y}\|/\| {\bf n}\|$ & $[1,10]$\\
  Fuzzy rules $R$ & 3\\
  Input features $D$ & $20=(K+1)3+2$\\
  {} & (elevation,azimuth,separation)$\times K$\\
  {} & log(frequency),stations\\
  Training samples & 1300\\
\end{tabular}
\end{table}
In Fig. \ref{data_dist}, we show the distribution of the various geometric parameters in the simulated data. The geometry of each observation is basically determined by the azimuth and elevation of each direction (including the target) and the angular separation of each outlier from the target. As seen from Fig. \ref{data_dist}, we have almost uniform coverage of the sky with these variables. Note also that the separation of the target from itself is always zero as seen in Fig. \ref{data_dist} (c), but we kept this redundant input to test the output of the learned fuzzy membership functions.

\begin{figure}[htbp]
  \begin{minipage}{0.99\linewidth}
\begin{center}
  \begin{minipage}{0.99\linewidth}
\centering
  \centerline{\includegraphics[width=0.6\textwidth]{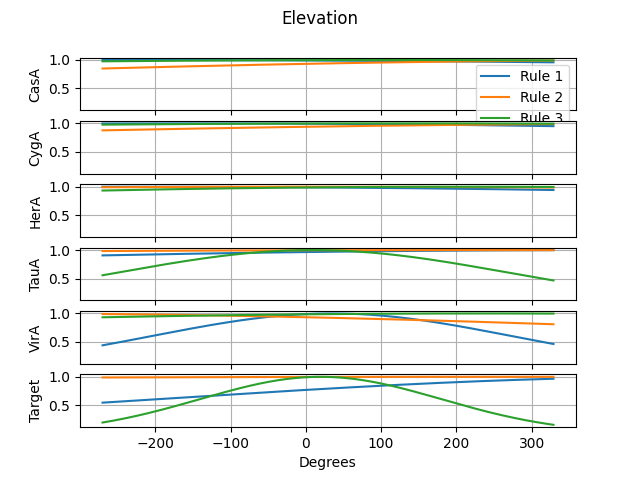}}
\vspace{0.01cm} \centerline{(a)}
\end{minipage}
  \begin{minipage}{0.99\linewidth}
\centering
  \centerline{\includegraphics[width=0.6\textwidth]{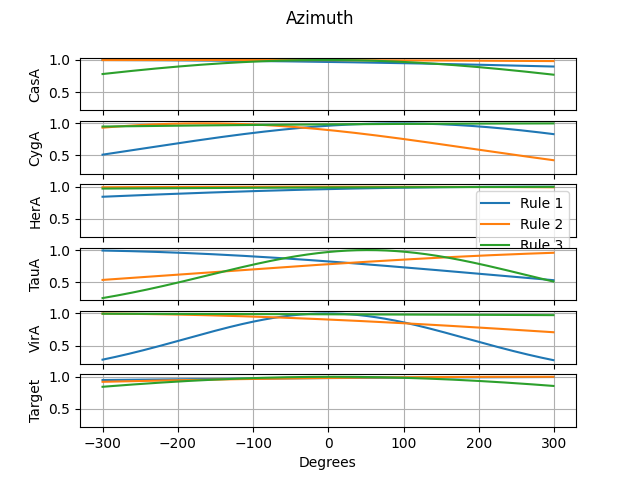}}
\vspace{0.01cm} \centerline{(b)}
\end{minipage}
  \begin{minipage}{0.98\linewidth}
\centering
  \centerline{\includegraphics[width=0.6\textwidth]{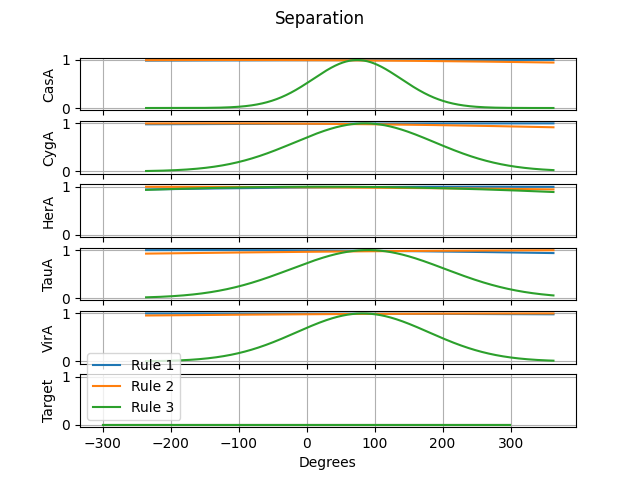}}
\vspace{0.01cm} \centerline{(c)}
\end{minipage}
\end{center}
\end{minipage}
  \caption{The learned Gaussian membership functions for (a) elevation (b) azimuth (c) separation. Looking at their complexity, we see that the azimuth plays an equally important role as the elevation in making a correct choice (unlike the separation, which is probably dependent on the azimuth and elevation). \label{fig_gmf}}
\end{figure}

\begin{figure}[htbp]
  \begin{minipage}{0.99\linewidth}
\begin{center}
  \begin{minipage}{0.99\linewidth}
\centering
  \centerline{\includegraphics[width=0.6\textwidth]{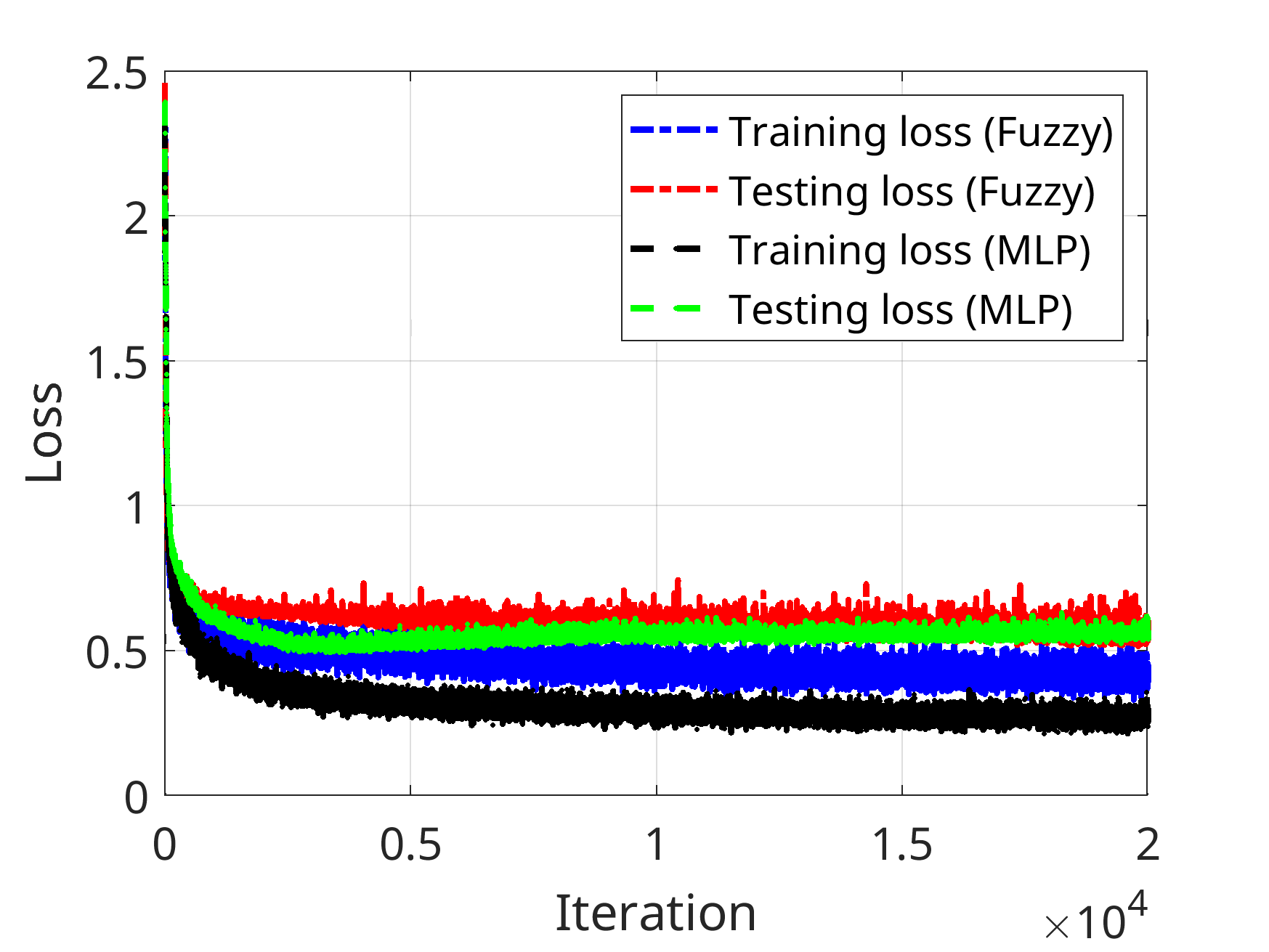}}
\vspace{0.1cm} \centerline{(a)}
\end{minipage}
  \begin{minipage}{0.98\linewidth}
\centering
  \centerline{\includegraphics[width=0.6\textwidth]{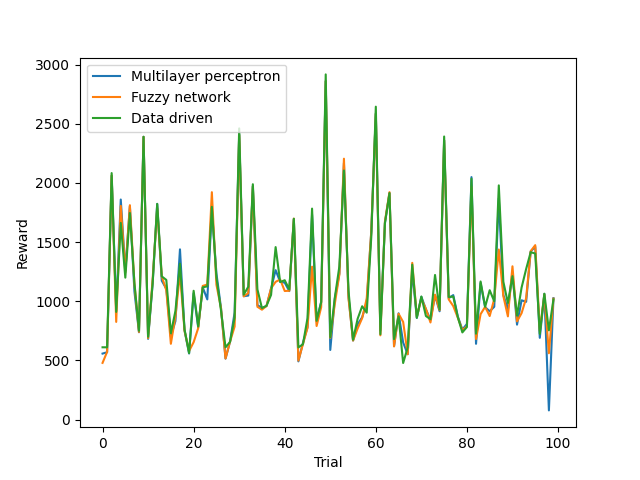}}
\vspace{0.1cm} \centerline{(b)}
\end{minipage}
\end{center}
\end{minipage}
  \caption{The training losses and the performance of the trained ML model compared with a purely data driven approach. (a) training/testing losses (b) reward (negative AIC) evaluation of the trained ML models compared to the data-driven approach. \label{training}}
\end{figure}

In order to build the deep neural network with a fuzzy input layer, we use the PyTSK toolkit \cite{PyTSK}. For comparison, we also trained a traditional multilayer perceptron using the same data. The outputs of both models are a $K\times 1$ vector of probabilities of each outlier being selected to be included in $\mathcal{I}$ of (\ref{cal}) ($>0.5$ being selected). The training and testing losses (mean squared error) for both models are shown in Fig. \ref{training} (a). The trained models are compared to the data driven approach in Fig. \ref{training} (b) using additional simulations. The reward for evaluation is directly calculated using the negative AIC (\ref{AIC}). We get close agreement in all three approaches in Fig. \ref{training} (b), but the data-driven approach performs an exhaustive search of all possible configurations of $\mathcal{I}$ and is therefore computationally expensive. The ML model-based approaches are much faster. Compared to the traditional multilayer perceptron based approach, the fuzzy inference based approach is more explainable as seen if Fig. \ref{fig_gmf}.

By looking at the learned Gaussian membership functions in Fig. \ref{fig_gmf}, we draw several conclusions. (i) Using the target separation (always zero) adds no extra information and can be removed from the input. (ii) Both azimuth and elevation has information that are used to produce the output (their membership functions have more variation). This contrasts with the use of separation as seen in Fig. \ref{fig_gmf}, which has only one membership function with variation (indicating only one active rule). (iii) The elevation membership functions for outliers that does not go below the horizon (CasA, CygA) has less variation than the other outliers, but HerA is an exception (probably due to lack of variation in the training data). 

To summarize, the increased explainability of the ML model enables us to: detect and exclude redundant input, detect examples that are rare and generate more data covering such cases, and detect more influential and also less influential input features. Further explainability can be added by using semantic rule mining techniques as future work \cite{Camastra2024}.

\section{Conclusions\label{sec:conclusions}}
We have proposed a model-based ML approach with added explainability for the determination of the optimal configuration of outlier removal in radio interferometric data. Simulated results show that the trained ML model is capable of performing comparable to a data-driven approach (and probably is better when the noise is high). We also show that by adding the fuzzy input layer, we can peer into the learned ML models and add further enhancements. Future work will expand the capabilities of such fuzzy inference and deep learning systems to handle more input parameters and other configuration parameters as output. 

\bibliographystyle{IEEE}
\bibliography{references}
\end{document}